\documentclass{aa}  
 
\usepackage[varg]{txfonts} 
\usepackage{graphicx} 
\usepackage{natbib} 
\bibpunct{(}{)}{;}{a}{}{,} 
 
\begin{document} 
 
   \title{Correlation of HI shells and CO clumps in the outer Milky Way} 
 
   \author{S. Ehlerov\'a 
          \and 
          J. Palou\v{s} 
          } 
 
   \institute{Astronomical Institute, Academy of Sciences, 
              Bo\v{c}n\'{\i} II 1401, Prague, Czech Republic\\ 
              \email{sona.ehlerova@asu.cas.cz} 
             } 
 
   \date{Received July 1, 2015; accepted December 3, 2015} 
 
   \abstract 
   {HI shells, which may be formed by the activity of young and massive  
stars, or connected to energy released by interactions of high-velocity  
clouds with the galactic disk, may be partly  
responsible both for the destruction of CO clouds and for the creation of others. 
It is not known which effect prevails.} 
   {We study the relation between HI shells and CO in the outer parts 
of the Milky Way, using HI and CO surveys and a catalogue of previously  
identified HI shells.} 
   {For each individual location, the distance to the nearest HI shell is 
calculated and it is specified whether it lies in the interior 
of an HI shell, in its walls, or outside an HI shell. The method takes 
into account irregular shapes of HI shells.} 
   {We find a lack of CO clouds in the interiors of HI shells and their 
increased occurrence in walls. Properties of clouds differ for different 
environments: interiors of HI shells, their walls, and unperturbed 
medium.}
   {CO clouds found in the interiors of HI shells are those that survived
and  were robbed of their more diffuse gas. Walls of HI shells 
have a high molecular content, indicative of an increased rate of 
 CO formation. Comparing the CO fractions within HI shells and outside  
in the unperturbed medium, we conclude that HI shells are responsible 
for a $\sim$ 20 \% increase in the total amount of CO in the outer  
Milky Way.} 
 
   \maketitle 
 
\section{Introduction}

The interstellar medium (ISM) is full of structures on all scales, from sub-pc to kpc. 
The larger structures, among which the HI shells, were discovered first; the shells were found in 
the HI distribution ranging in sizes from a few pc to about 1 kpc. The first list of 
HI shells in the Milky Way was published by  
\citet{1979ApJ...229..533H} and others followed: 
\citet{2002ApJ...578..176M}, \citet{2005A&A...437..101E}, 
\citet{2013A&A...550A..23E}, \citet{2014A&A...564A.116S}. 
HI shells were also identified in other galaxies  
\citep[for a review on shells in external galaxies 
and an analysis of HI shells in THINGS galaxies see][]{2011AJ....141...23B}. 
 
Since early days it has been speculated that these shells are the result of energetic activities 
of massive stars: winds, radiation, and supernova explosions. For many large shells, the energy 
needed to create them must have come from the whole cluster of stars, from OB associations. 
Proving   this connection is not completely straightforward since many shells are older than 
the expected lifetime of massive stars and the results are often confusing 
\citep{1999AJ....118..323R,2000ApJ...529..201S}. 
However, the statistical correspondence between  HI shells and stars 
\citep{2013A&A...550A..23E, 2014A&A...564A.116S} indicates
the relation.  
 
Many observational papers, both for the Milky Way and external galaxies, 
show that HI shells frequently exist at large galactocentric distances, far from star 
forming regions and often with quite large sizes. This implies, that other mechanisms 
might be employed to explain these structures: ram pressure \citep{2002AJ....123.1316B} 
or the infall of a high-velocity cloud to the disk \citep{1987A&A...179..219T}. 
These events might be responsible for a fraction of HI shells, but probably not 
for the majority. 
 
HI shells evolve in a gaseous disk. They are sensitive to local density distribution, 
but since they are quite big, they quickly overgrow the smaller pc scale fluctuations. 
Once their dimensions are  comparable to the disk scale --- which is not unusual --- 
their shapes should be influenced by the large-scale  density gradient  
in the disk and they may prolong in the z-direction (these prolonged shells are usually called 
worms). If the interior of the shell is still hot 
--- i.e. if the progenitor massive stars still exist --- 
this hot gas may flow into the galactic halo (the worms become chimneys).  
In such a way HI shells may influence the energetic flows in galaxies. 
For dwarf galaxies with low gravity \citep[e.g.][]{2009A&A...493..511V} 
or for starburst galaxies (e.g. M82), such an event could mean the loss of 
this hot gas to the intergalactic medium. 
 
With the advent of infrared satellites, structures with pc sizes were discovered in 
the dust emission. They were found to be almost everywhere in the disk and therefore 
we talk about the ``bubbling galactic disk''  
\citep{2006ApJ...649..759C, 2010A&A...523A...6D, 2012MNRAS.424.2442S}. 
However, the infrared bubbles are quite different from HI shells. 
Some of them are connected to young stars; basically, each massive star is able 
to produce an HII region, which is surrounded by the photodissociation region 
identified as the infrared bubble \citep{2012A&A...542A..10A}.  
They are often found in the vicinity of CO clouds and other indicators of the on-going 
star formation. One might expect, that some of these bubbles --- or mergers of several 
of these bubbles --- may eventually grow to become HI shells. 
The bubbles discovered in the MIPSGAL 24 $\mu$ survey  
\citep{2010AJ....139.1542M} may have different origins; some of them are connected to 
planetary nebulae and late phases of stellar evolution.  
  
Some of the infrared bubbles, as already stated, are directly  connected to the 
on-going (or more correctly, very recent) star formation. It is speculated 
that the bubbles might trigger the new star formation as their walls are the homes of objects younger 
than the bubble itself  \citep[][and many more]{2015MNRAS.452.2794W, 2014A&A...565A...6S}, 
either by the collect \& collapse scenario or by  radiation driven implosion. Another 
interpretation might be that they are only the redistributors of the star formation 
that would happen anyway. 
 
For HI shells, the situation is even less clear. They are --- or were --- strong shocks 
\citep{1988ARA&A..26..145T} and as such we can imagine for them both the inhibiting 
model, i.e. the destruction of existing molecular clouds by  intense radiation, strong 
winds, and supernova explosions, or the supporting/stimulating model, i.e. the increased 
cloud formation in the swept-up gas around clusters of young stars. 
 
For small dwarf galaxies the star formation taking place mostly in the centre of the 
galaxy may create one supergiant HI shell, which then significantly influences the 
secondary star formation 
\citep[as an example see the Holmberg I galaxy, ][]{2001AJ....122.3070O}. 
However, the Milky Way is a large galaxy that has more complex patterns of  star formation.

\citet[][]{2013ApJ...763...56D} studied the relationship between HI and CO in 
supergiant shells in the Large Magellanic Cloud (LMC). They found out that $\sim$ (4-10)\% 
of the total molecular mass of the LMC was created as a by-product of the stellar 
activity that formed the shells. \citet[][]{2011ApJ...728..127D} studied two supershells 
in the Milky Way and found that both objects show an enhanced level of molecularization. 
Both these studies indicate the stimulating effect of supershells on the molecularization 
and the total amount of CO in galaxies. 
 
Unlike the two papers mentioned,  we want to study the effect of all shells, 
not only supershells, which are striking and important examples of an 
HI shell population, but are not  typical. The origin of the shells  
and supershells may be different, and the net outcome of all shells, small and large, 
could differ from the outcome of supershells. We analyse the relation between CO and HI 
bubbles; more specifically, we discuss whether CO is concentrated in the walls of HI bubbles, whether there is little or no  correlation between CO and HI bubbles, and how 
the position of CO in relation to HI shells influences the CO properties.

\section{Data} 
 
In our analysis we use two kinds of input datacubes: HI data and CO data, and also a 
catalogue of HI shells by \citet[][Paper 1]{2013A&A...550A..23E}. We focus on the outer 
parts of the Galaxy and  avoid the  inner parts, which are too full of HI shells and 
bubbles.  Owing to the high filling factor of bubbles, in the inner Galaxy it is difficult to 
define their shapes precisely (or at least reasonably precisely;  see the discussion 
in Paper 1). Therefore, we restrict our study to the strip 
$b \in (-5^{\circ},+5^{\circ})$ in the 2nd and 3rd Galactic quadrants: $l \in (90^{\circ}, 270^{\circ})$. 
 
The longitude-latitude-velocity (lbv) datacubes contain a lot of ``empty space'';  
the signal is 
below the sensitivity limit because there is too little gas at a given location 
and velocity  because the corresponding 
velocities are forbidden (for each quadrant about half of the cube lies at 
forbidden velocities),  because there is no gas there (e.g. because it is too far from the 
Galactic centre), or because the sensitivity of the survey is simply too low. 
To avoid the empty lbv regions, we always study only pixels/places with some HI 
emission $T_{\mathrm{HI}} \geq T_{\mathrm{HIcutoff}}$ (see the next section for the 
choice of $T_{\mathrm{HIcutoff}}$).  
 
\begin{figure*} 
\centering 
\includegraphics[angle=-90,width=0.9\linewidth]{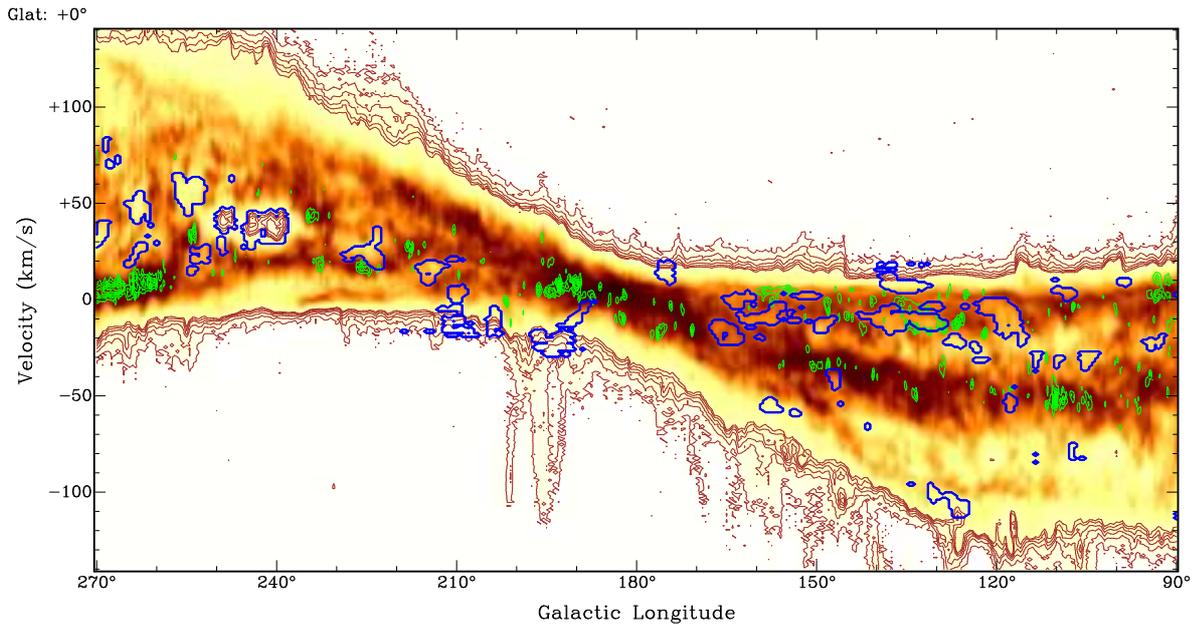} 
    \caption{Galactic longitude-velocity map at the Galactic latitude $b = 0^{\circ}$ 
     of the outer Milky Way: HI (map), CO (green contours) and HI shells (blue outlines). 
     } 
       \label{lv_b_HICObub} 
\end{figure*} 
 
\subsection{HI data} 
 
We use the Leiden/Argentina/Bonn HI survey \citep[LAB;][]{2005A&A...440..775K}. 
It is an all-sky survey, a combination of observations made from two instruments. The 
angular resolution (HPBW) is $\sim 0\fdg6$, the velocity resolution is 
$1.3\ \mathrm{kms}^{-1}$, and the LSR velocity covers the interval
$(-450,450)\ \mathrm{kms}^{-1}$.  
The HI datacube contains the brightness temperature; the rms noise of the LAB  
survey is (0.07-0.09) K.

The pixel size of the LAB datacube is $0.5^{\circ}$ and the channelwidth is 
$\Delta v = 1.0306\ \mathrm{kms^{-1}}$. As mentioned above, to avoid the empty regions 
of the lbv we have to define a certain cutoff of an HI emission $T_{\mathrm{HIcutoff}}$. 
We choose the HI cutoff to be 0.3 K, which equals the 3 $\sigma$ level of noise in 
the HI data, and which also corresponds to values chosen for the detection of 
HI shells (see Paper I). It corresponds to a column density of 
$5.5\  10^{17}$ cm$^{-2}$.   
 
We performed some calculations with other HI cutoffs to test the dependence of our results 
on this value. If not  otherwise stated, when we refer to `the studied datacube' it means 
that we restrict ourselves to the parts of it that have $T_{\mathrm{HI}} \geq 0.3\ K$. 
 
\subsection{CO data} 
 
We use the CO  (J = 1->0) survey of \citet[][]{2001ApJ...547..792D}. It is 
a composite survey of the Milky Way, consisting of observations from several telescopes. 
It is not an all-sky survey, but it contains the whole Galactic plane and some  additional regions. 
Resolutions, samplings, and sensivities of the different surveys vary. 
 
The combined datacube  (the so-called deep CO survey) that we use has a pixel 
size of $0.125^{\circ}$ and a channelwidth of $\Delta v = 1.3\ \mathrm{kms^{-1}}$.  
In the 2nd and 3rd quadrants it fully covers the Galactic plane region in the 
belt
$b \in (-5^{\circ},+5^{\circ})$. 
 
The datacube gives the main beam temperature;  the rms noise is 0.1 K.  
However, probably due to the `drastic noise reduction' mentioned in 
\citet[][]{2001ApJ...547..792D}, it is safer to believe only the data above $\sim 5\sigma$.  
  
\subsection{HI shells} 
  
As an input catalogue of HI shells we use our own catalogue from Paper 1. Shells in 
this list were identified automatically as continuous regions of a lower HI temperature 
surrounded by a higher temperature wall or region. The search was done in 2D lb-maps and 
then  the detected low-temperature regions were combined into 3D lbv structures. 
There were no prescriptions of the shape of the HI shell, but the structures 
had to span at least eight consecutive velocity channels (i.e. $8\ \mathrm{kms}^{-1}$), 
and their spectra had to show a temperature contrast of at least 4 K towards their 
surroundings. The overlap between 2D lb-cuts of the 3D lbv-structure  
identified in the subsequent channel maps must be large enough and they should have 
a similar size and shape (see Paper I for more precise discussion). 
The minimum angular dimension of the detected structure is $1\fdg5$, the maximum is 
$45^{\circ}$. The algorithm cannot detect entirely open structures such as very 
evolved galactic chimneys or structures with completely fragmented walls (see also
the discusion in  Section 5.3). 
Identified structures are low-temperature regions, i.e. interiors of HI
shells. Walls are not detected by our algorithm and  the dimensions used here
and taken from Paper 1 also refer only to sizes of the interiors.
 
The analysis of the HI LAB survey \citep[LAB;][]{2005A&A...440..775K} using methods and 
conditions briefly summarized in the paragraph above we identified 333 shells. By looking 
at results and by comparing our findings with other catalogues of HI shells  
\citep{2002ApJ...578..176M,2014A&A...564A.116S} we found that the identifications 
in the outer Milky Way are more reliable than identifications in the inner Milky Way, 
probably owing to a heavy overlapping of shells in the inner disk. This is the main 
reason, why in the current study we are only interested in the outer Milky Way. 
 
The studied datacube contains the whole volume or a significant fraction of the volume
of 70 HI shells from the catalogue plus some smaller parts of ten more shells. For the 
current analysis we use the real 3D pixel maps of shells; the idealized dimensions 
($\Delta l_0$, $\Delta b_0$, $\Delta v_0$) and positions ($l_0$, $b_0$, $v_0$) of these 
shells are given in Table 1 of Paper I.  As an example Fig. \ref{lv_b_HICObub} 
shows the HI lv-map at ($b = 0^{\circ}$) with CO contours and HI shells outlines 
overlaid. HI shells do not overlap significantly in the studied datacube 
and we see that CO avoids the internal volume of shells.  
 
\subsection{Comparing HI and CO datacubes} 

Pixel sizes and channelwidths of the HI and CO datacubes differ. Therefore,  
we construct an auxiliary grid with high resolution and for each pixel of this  
auxiliary grid we find the closest HI or CO pixel and their appropriate values.  
All grid calculations are done in this auxiliary grid. We do not interpolate  
the values or regrid the HI or CO datacubes, but we keep their original resolution.  
 
This is only one possible method; we made a few calculations 
with datacubes regridded to the lower resolution of the two original datacubes.  
The results were similar to the results of our current approach.

\section{Methods} 
 
Here we describe how we attribute the HI shells and their parts to studied pixels, and 
which quantities we are interested in. 
 
\subsection{Composition of an HI shell} 
 
An HI shell consists of an HI bubble,  sometimes also called a hole, and an HI wall. 
A bubble is the interior of the shell, a region of  lower HI temperature (i.e.  density). 
This structure was identified by the searching algorithm in Paper 1 (and it is an 
input quantity). The wall is, from the technical point of view, a region just neighbouring 
the bubble. 
 
\subsection{Different types of pixels} 
 
For each pixel of the studied datacube we have to know  
two things: first, whether it is inside the HI bubble or not 
and second, how far it is from the wall of the bubble (if 
it lies inside it) or how far it is from the wall of the nearest  
HI bubble (if it is outside).   We know the first from the HI  
shells datacube (section {\it HI shells});  we have 
to calculate the second from this datacube. 
 
For pixels that do not lie inside any HI bubble, we have to find  
the nearest HI bubble. For each such outside pixel P1 we 
calculate its distance to all pixels that are inside  
HI bubbles, and the final distance of the P1 
pixel is the minimum from these calculated distances. 
  
We do not use the absolute value of the distance (i.e. in units of pixels) 
but the relative distance, which depends on dimensions of the bubble 
towards which the distance is calculated, 
\begin{equation} 
{r_{\mathrm{rel}}} = 
\sqrt{\left({l-l_{nearest}} \over {0.5\Delta l_0}\right)^2 + 
\left({b-b_{nearest}} \over {0.5\Delta b_0}\right)^2 + 
\left({v-v_{nearest}} \over {0.5\Delta v_0}\right)^2} 
\label{eq:rreldef} 
,\end{equation} 
where $\Delta l_0$, $\Delta b_0$, $\Delta v_0$ are 
dimensions of the bubble (Table 1 in Paper 1), coordinates of 
the outside pixel are $l,b,v$ and coordinates of the inside  
pixel are $l_{nearest}$, $b_{nearest}$, $v_{nearest}$. 
As we are dealing only with pixels near the equator,  
Eq. \ref{eq:rreldef} is precise enough even without the 
$cos(b)$ correction. 
We intentionally use the distance towards the nearest pixel 
of the HI bubble and not  the distance towards the centre 
of the shell, for example. Our choice takes  irregular 
shapes of HI holes  into account better. It is also less sensitive 
to inaccurate measurements of shell dimensions and positions. 
 
The choice of the relative distance has some advantages over 
using the absolute distances. First, we remove the dependence 
on units (degrees and $kms^{-1}$). Second, the difference between  
angularly different shells is suppressed (the same physical 
distance between two objects depends on the distance of the 
observer and by using the ratio between the projected distance 
and dimensions we take this into account). The same 
holds for the offset in the LSR velocity. 
 
We are dealing with expanding structures 
and therefore we cannot convert the lbv-datacube into an xyz-datacube  
using kinematic distances;  it would be too inaccurate or 
simply wrong. For the whole structures (HI shells, CO clumps) 
we can calculate the distance and physical dimensions, but 
for the work with all pixels in the datacube 
we are limited to the lbv-space and our relative distance. 
 
The value of $r_{\mathrm{rel}}$ is $ > 0$ for pixels outside HI bubbles. 
 
For pixels inside HI bubbles we calculate the relative distance 
to the  centre of the shell. In Eq. \ref{eq:rreldef}, instead 
of ($l_{nearest}$, $b_{nearest}$, $v_{nearest}$), we take coordinates of the shell 
centre given in Table 1 of  Paper I, and then subtract 1. It makes 
$r_{\mathrm{rel}} < 0$: pixels close to 
the wall have $r_{\mathrm{rel}}$ close to 0, while those 
close to the centre of the bubble have $r_{\mathrm{rel}}$ 
close to -1.  
 
According to its relative distance $r_{\mathrm{rel}}$ we ascribe 
each pixel its type: 
\begin{enumerate} 
\item{{\it inner bubble:} inside one of the HI bubbles with   
       $r_{\mathrm{rel}} \leq -0.25$} 
\item{\ {\it outer bubble:} inside one of the HI bubbles with   
       $0 > r_{\mathrm{rel}} > -0.25$} 
\item{{\it inner wall:} not part of any HI bubble with    
       $0 < r_{\mathrm{rel}} \leq 0.2$} 
\item{{\it outer wall:} not part of any HI bubble with    
       $0.2 < r_{\mathrm{rel}} \leq 0.5$} 
\item{{\it outside wall:} not part of any HI shell with  
       $0.5 < r_{\mathrm{rel}} \leq 1.0$} 
\item{{\it far outside:} not part of any HI shell with   
       $1.0 < r_{\mathrm{rel}} \leq 2.0$} 
\item{{\it unperturbed:} not part of any HI shell with   
       $r_{\mathrm{rel}} > 2.0$} 
\end{enumerate} 
These terms are simply names. These names could be  
Environment 1, 2, etc., but 
we believe that our designation is slightly more illustrative. 
The division between different types was made
by looking at profiles of many structures (see below) and by   
dividing the studied pixels into groups of roughly the same 
size. The interior (i.e. the bubble) was divided into two parts, inner 
and outer, to see whether there is any difference in behaviour  
between parts of the bubble close to walls and further from them.  
The same holds for walls. 
 
\subsection{$T_{\mathrm{HI}}$ and CO filling factor and 
the molecularization level} 
 
The good way to describe HI in some types of environments 
(e.g. inside one of the HI bubbles, see the previous section for the list 
of environments) is the  average brightness temperature $T_{\mathrm{HI}}$, 
which is an estimate of the density in the given type of  environment.  
However, for the molecular medium it is not as good 
because the medium is clumpy and often optically thick, 
and temperatures in CO clumps are not 
the straightforward description of densities in the medium.  
Instead, we calculate 
the CO filling factor $f_{\mathrm{CO}}$ of a region or pixel type 
as the ratio of a number of pixels with a significant CO temperature 
to a total number of pixels. `Significant CO temperature' means 
$T_{\mathrm{CO}} \geq T_{\mathrm{COcutoff}}$. 
 
The level of molecularization --- the ratio of the 
molecular density to the total density --- 
in a given place/pixel with 
observed HI and CO temperatures is given by $\varepsilon$ 
\begin{equation} 
\varepsilon = {{T_{\mathrm{CO}} X\Delta v_{\mathrm{CO}}} \over 
              {T_{\mathrm{CO}} X\Delta v_{\mathrm{CO}} + 
               T_{\mathrm{HI}} A\Delta v_{\mathrm{HI}}}} 
            = {{T_{\mathrm{CO}}} \over {T_{\mathrm{CO}} + \eta T_{\mathrm{HI}}}} 
\label{eq:epsdef} 
,\end{equation} 
where $\Delta v_{\mathrm{CO}}$ and $\Delta v_{\mathrm{HI}}$ are the 
channel widths of CO and HI datacubes (which are similar, but not the 
same), $X$ is the famous X factor for which we use the value 
$X=1.8\times10^{20} cm^{-2}K^{-1}(kms^{-1})^{-1}$, and 
$A=1.82\times10^{18} cm^{-2}K^{-1}(kms^{-1})^{-1}$. The value of $\eta$ is 
\begin{equation} 
\eta = {{A\Delta v_{\mathrm{HI}}} \over {X\Delta v_{\mathrm{CO}}}} 
     \simeq 0.008 
\label{eq:etadef} 
.\end{equation} 
 
CO is often optically thick and then using the simple relation between 
the density and brightness (main-beam) temperature underestimates 
the real density in the environment. In the case of the optically 
thick medium, $\varepsilon$ from Eq. \ref{eq:epsdef} is artificially 
lower than its real (unknown) value.  
 
By using the CO cutoff temperature, it can be understood  that we are interested 
in the regions  where the particle density of the molecular medium 
forms at least a fraction $\varepsilon_{\mathrm{sigCO}}$ of the 
total density (Eq. \ref{eq:epsdef}), 
$$ \varepsilon_{\mathrm{sigCO}}  = {{T_{\mathrm{COcutoff}}} \over {T_{\mathrm{COcutoff}} + \eta T_{\mathrm{HImax}}}}, $$ 
where $T_{\mathrm{HImax}}$ is the maximum HI temperature (150 K in the  
LAB survey for the 2nd and 3rd quadrants). 
Quantified for the $T_{\mathrm{COcutoff}} = 0.6\ K$ it means that 
the molecular medium forms at least one third of the total particle density 
in the region with a significant CO emission. 
Translated into masses it says  that in such regions at least 2/3 of the 
mass is molecular. 
 
\subsection{Relative profiles} 
 
Using $r_{\mathrm{rel}}$ from Eq. \ref{eq:rreldef} we calculate the 
average profiles of $T_{\mathrm{HI}}$, $f_{\mathrm{CO}}$, and  $\varepsilon$ 
{as the average value for a given interval of $r_{\mathrm{rel}}$.} 
We can do it for the whole studied area, i.e.  the 2nd and 3rd 
quadrant, or for each HI shell separately. Using the relative distance 
$r_{\mathrm{rel}}$ has the advantage of scaling the distances of pixels  
to the dimension of a given shell.  
 
The average profile, composed of profiles of all shells, is 
dominated by angularly large structures since the relative weight 
of the shell in the composite picture depends on the amount of  
pixels the shell contains. 
 
\subsection{CO clumps} 
 
\begin{figure} 
\centering 
\includegraphics[angle=-90,width=0.9\linewidth]{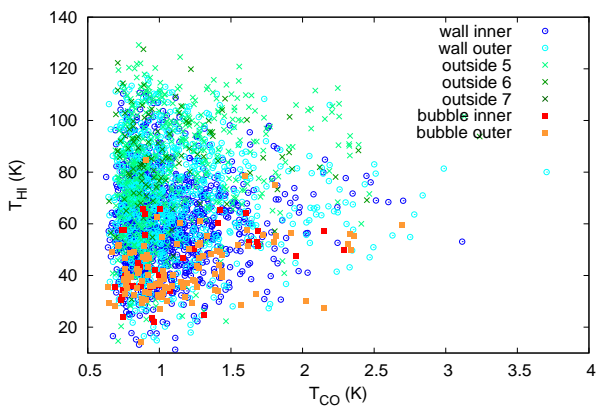} 
    \caption{Relation between CO and HI temperatures 
      ($T_{\mathrm{cl,CO}}$, $T_{\mathrm{cl,HI}}$) of CO clumps.} 
       \label{cl_tcothi} 
\end{figure} 
 
Another way to describe the CO distribution is to use the natural 
clumpiness of CO and deal with CO clumps, not with the smooth  
continuous distribution. We identify clumps in the CO using the DENDROFIND  
algorithm described by \citet[][]{2012A&A...539A.116W}. 
 
The clump-finding algorithm DENDROFIND is similar to the well-known  
CLUMPFIND (Williams et al. 1994), over which it has some advantages, mostly 
of the technical character. It searches for clumps that    
1)  are contiguous volumes in the position-position-velocity space; 
2) consist of more than a prescribed number of pixels ($Npxmin$); and 
3) where the difference between the clump peak temperature and the temperature 
at which it borders with the neighbouring clump is higher than a 
value $dTleaf$.  
The clump finding is performed only for pixels with a temperature 
above a cutoff $Tcutoff$. The fourth parameter of the DENDROFIND 
is the number of levels $Nlevels$ into which the temperature scale between 
the maximum temperature in the analysed datacube and the cutoff temperature 
$Tcutoff$ is divided;  for large values of $Nlevels$, the results of DENDROFIND 
cease to depend on it. Overall, results of DENDROFIND do depend on technical 
parameters ($Npxmin$, $dTleaf$, and $Tcutoff$), but the dependence is 
predictable and not steep. 
 
With the cutoff temperature $Tcutoff = T_{\mathrm{COcutoff}}$  
set to 0.6 K, the minimum size of the clump set to $Npxmin = 5$ and 
the difference between clumps $dTleaf = 0.3 K$,  
DENDROFIND analysis produces 2617 clumps.  
The number  changes slightly when using slightly different values 
for parameters $Npx$ and $dTleaf$  and  changes significantly for 
different cutoff temperatures $T_{\mathrm{COcutoff}}$ (see   
section 4). 
 
For further analysis each clump is characterized by coordinates of 
its centre of mass, its radius, average HI temperature $T_{\mathrm{cl,HI}}$, 
its maximum and average CO temperatures  
$T_{\mathrm{max,cl,CO}}$ and $T_{\mathrm{cl,CO}}$, average molecularization  
$\varepsilon_{\mathrm{cl}}$,  and  number of pixels. Radius of a clump is  
defined as the average distance of pixels belonging to it  from its centre. 
We also calculate the average HI temperature just outside the clump 
(e.g. from its edge to a distance equal to three times clump radius  
from its centre  
$T_{\mathrm{cl,HI, out}}$. Fig. \ref{cl_tcothi} shows  
the average HI and CO temperatures for identified clumps.

\section{Results} 
 
\subsection{CO in HI shells} 

\begin{figure*} 
\centering 
\includegraphics[angle=-90,width=0.45\linewidth]{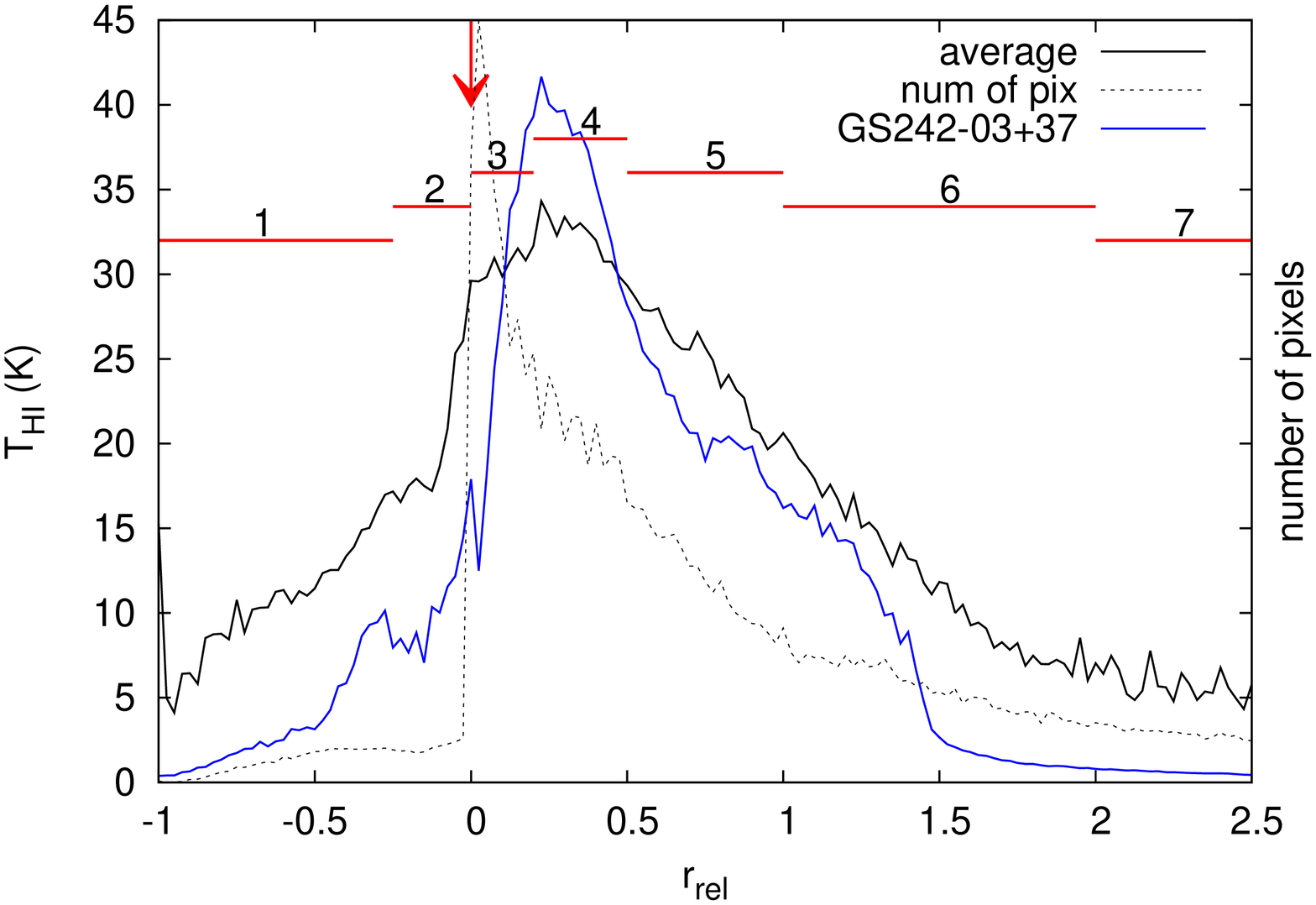} 
\includegraphics[angle=-90,width=0.45\linewidth]{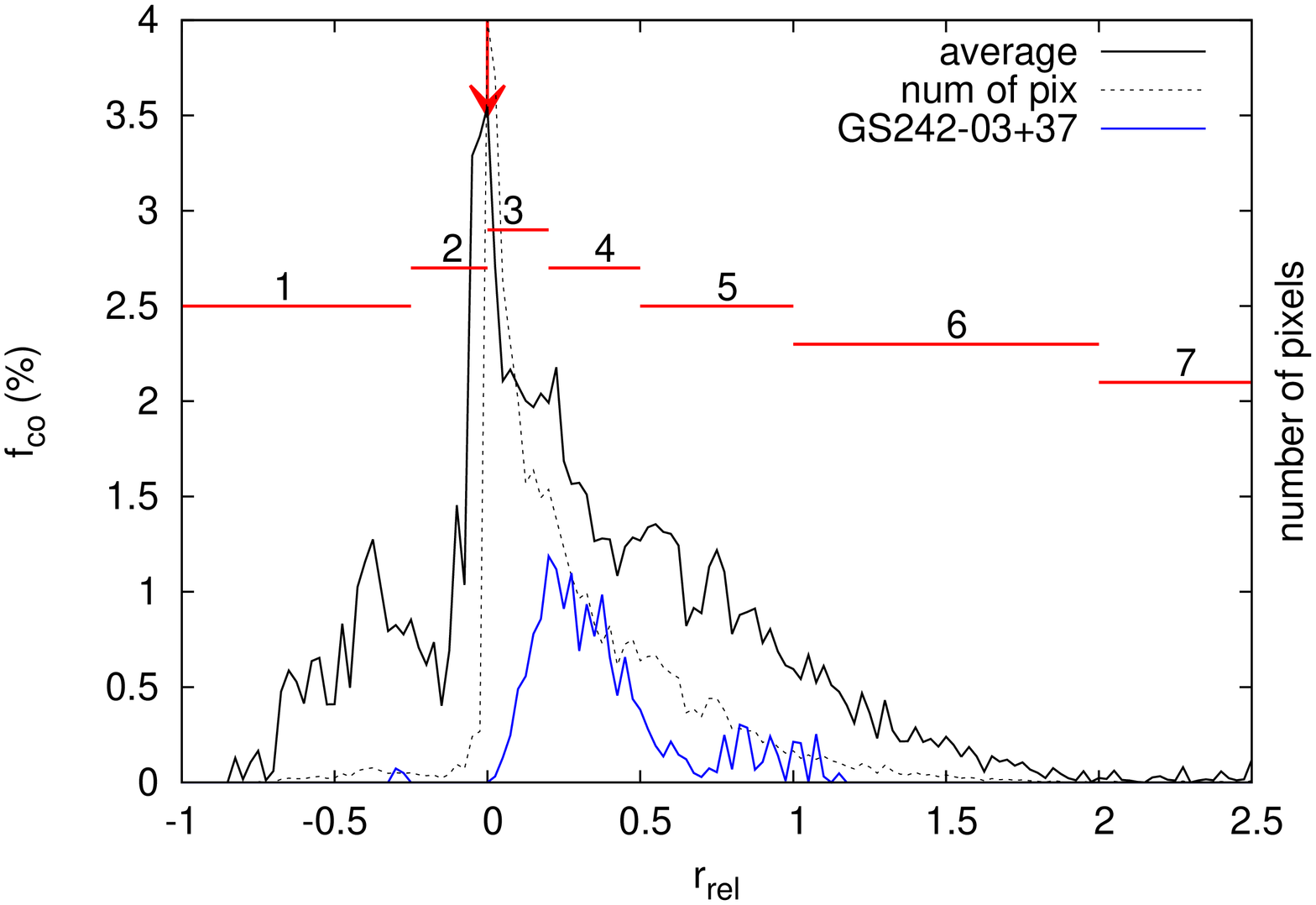} 
    \caption{The average HI temperature (left) and the CO filling factor 
(right) as a function of the relative distance $r_{\mathrm{rel}}$. Black 
solid line shows the average for the whole datacube, blue solid 
line shows the profile for one Galactic HI supershell, GS242-03+37. 
Dotted lines show the number of pixels, from which the average was calculated. 
Red lines with numbers give extents of different 'types of environment'. 
Red arrows indicate the transition between bubbles and walls.} 
       \label{glob_rrel1} 
\end{figure*} 
 
\begin{table*} 
\caption{Different environments in the Galaxy: filling factors 
and CO clump properties. 
} 
\label{tab_env1}      
\centering                          
\begin{tabular}{|lr|rr|rrr|}        
\hline 
env & $q_{\mathrm{lbv}}$~~~ & 
$T_{\mathrm{HI}}$ & $f_{\mathrm{CO}}$ & $T_{\mathrm{COsig,HI}}$ & 
$T_{\mathrm{COsig,CO}}$ & $\varepsilon$~~~~ \\ 
\hline 
0:{\tiny all} & 1.000 & 22 & 1.13 & 68 & 1.30 & 0.68 \\ 
1:{\tiny inner bubble} & 0.028 & 12 & 0.67 & 46 & 1.42 & 0.76 \\ 
2:{\tiny outer bubble} & 0.026 & 22 & 1.99 & 40 & 1.45 & 0.79 \\ 
3:{\tiny inner wall} & 0.209 & 30 & 2.34 & 60 & 1.26 & 0.70 \\ 
4:{\tiny outer wall} & 0.204 & 32 & 1.49 & 72 & 1.30 & 0.66 \\ 
5:{\tiny outside} & 0.200 & 25 & 1.06 & 86 & 1.38 & 0.63 \\ 
6:{\tiny outside} & 0.182 & 13 & 0.28 & 85 & 1.29 & 0.63 \\ 
7:{\tiny outside} & 0.150 &  4 & 0.02 & 38 & 0.77 & 0.79 \\ 
\hline 
\end{tabular} 
\tablefoot{ 
The first part gives the type of the environment and its fraction 
in the whole datacube. The second part gives the average HI temperature 
in the environment and the CO filling factor (in \%). The third part gives 
average HI and CO temperatures in pixels with the significant CO emission 
in the environment and their average molecularization. 
} 
\end{table*} 
 
Table \ref{tab_env1} shows the type of the environment (first column) 
and its fraction $q_{\mathrm{lbv}}$ 
in the complete datacube (second column). 
Numbers label pixels of different environments:  
1 and 2  inside bubbles; 3 and 4 inside 
walls; 5, 6, and 7 outside shells; 0  for all environments (see 
  Methods  for a more detailed explanation). 
The  average HI temperatures $T_{\mathrm{HI}}$ are in the third column; 
the fourth column shows the CO filling factors $f_{\mathrm{CO}}$ 
(in \%), which tell how much space of the environment is occupied 
by regions with a significant CO emission, i.e. 
$T_{\mathrm{CO}} \geq 0.6\ K$, which is 
the same value as that used in DENDROFIND for the clump identification.   
The fifth, sixth, and seventh columns give average results for regions 
with significant CO emission only, i.e. inside CO clumps: 
the average HI and CO temperatures,  and molecularization 
($T_{\mathrm{COsig,HI}}$, $T_{\mathrm{COsig,CO}}$, $\varepsilon$). 
 
Bubbles themselves fill only a small fraction of the studied volume,   
walls occupy a much larger space. This is  a consequence of several  
factors: first, some of the  HI shells are not contained  
(or not completely contained) in the 
studied datacubes, but their walls are there. Furthermore, shapes of HI shells 
are not ideal/smooth/spherical and all higher density protrusions or 
even embedded structures with higher $T_{\mathrm{HI}}$ count as walls in 
the searching mechanism. 
 
Results in  Table \ref{tab_env1} depend on several quantities. 
The most important is the catalogue of shells, more precisely, their 
precise identification, as it gives the distribution of studied pixels 
into `environments' (Column 1). Then the results depend on cutoffs, 
especially on the CO cutoff (e.g. different values of the limit for  
the significant CO emission lead to different values of $f_{\mathrm{CO}}$), 
but also on the HI cutoff (though only 
mildly). However, the trends we observe, for example the dependence of 
$\varepsilon$ on the environment, are the same. 
 
Fig. \ref{glob_rrel1} shows the average profile of the $T_{\mathrm{HI}}$ 
and $f_{\mathrm{CO}}$ as a function of the relative distance $r_{\mathrm{rel}}$. 
It is the average of the whole datacube. 
Since HI bubbles were identified as local minima, it is not surprising 
that the average HI temperature is low inside them, with the HI temperature 
in the inner bubble (environment 1) being lower than in the outer bubble 
( environment 2). Maximum values are reached in walls, 
especially in outer walls (environment 4), with the distance  
$(1.2-1.5)r_{\mathrm{rel}}$ 
from the centre of the shell (or $(0.2-0.5)r_{\mathrm{rel}}$ from the transition 
bubble-wall). 
All these values have high dispersion and depend on the local conditions 
and on the position in the Galaxy. 
Fig. \ref{glob_rrel1} also shows the relative profile of one 
individual shell. It is the Galactic supershell GS242-03+37, one  
of the largest shells in the outer Milky Way and a very pronounced 
one. It was first identified by \citet{1979ApJ...229..533H}.  
  
How much of the space is occupied by the CO emission with 
($T_{\mathrm{CO}} \geq 0.6\ K$)?  
The  largest amount of CO is found in the inner walls of HI 
shells, more than in the outer wall, even though it is still 
quite a high value. A high value of $f_{\mathrm{CO}}$ is also found  in the 
outer bubble environment, close to the wall. It seems that  CO 
is more  concentrated than HI  towards the inner wall of the HI shell  
or towards the transition bubble-inner wall. 
 
The CO emission is always associated with high (or higher) HI temperatures, 
as seen in the fifth column of  Table \ref{tab_env1}.  
These values are not simply correlated to the average HI temperature  
in the environment. 
The average CO temperatures in regions with  significant 
CO emission (as  given in the sixth column)  
are slightly higher in bubbles than in walls. The 
molecularization behaves 
the other way round: $\varepsilon$ is higher inside bubbles than in walls.

\begin{figure} 
\centering 
\includegraphics[angle=-90,width=0.9\linewidth]{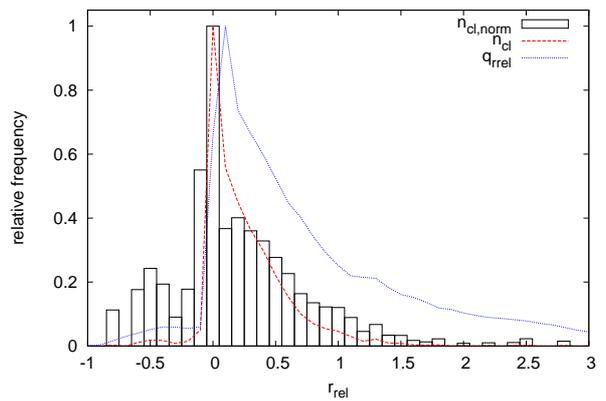} 
    \caption{Number of CO clumps as a function of their 
     relative distance $T_{\mathrm{CO}}$.} 
       \label{cl_hist_rrel} 
\end{figure} 
 
\subsection{CO clumps relative to HI shells} 

\begin{figure*} 
\centering 
\includegraphics[angle=-90,width=0.45\linewidth]{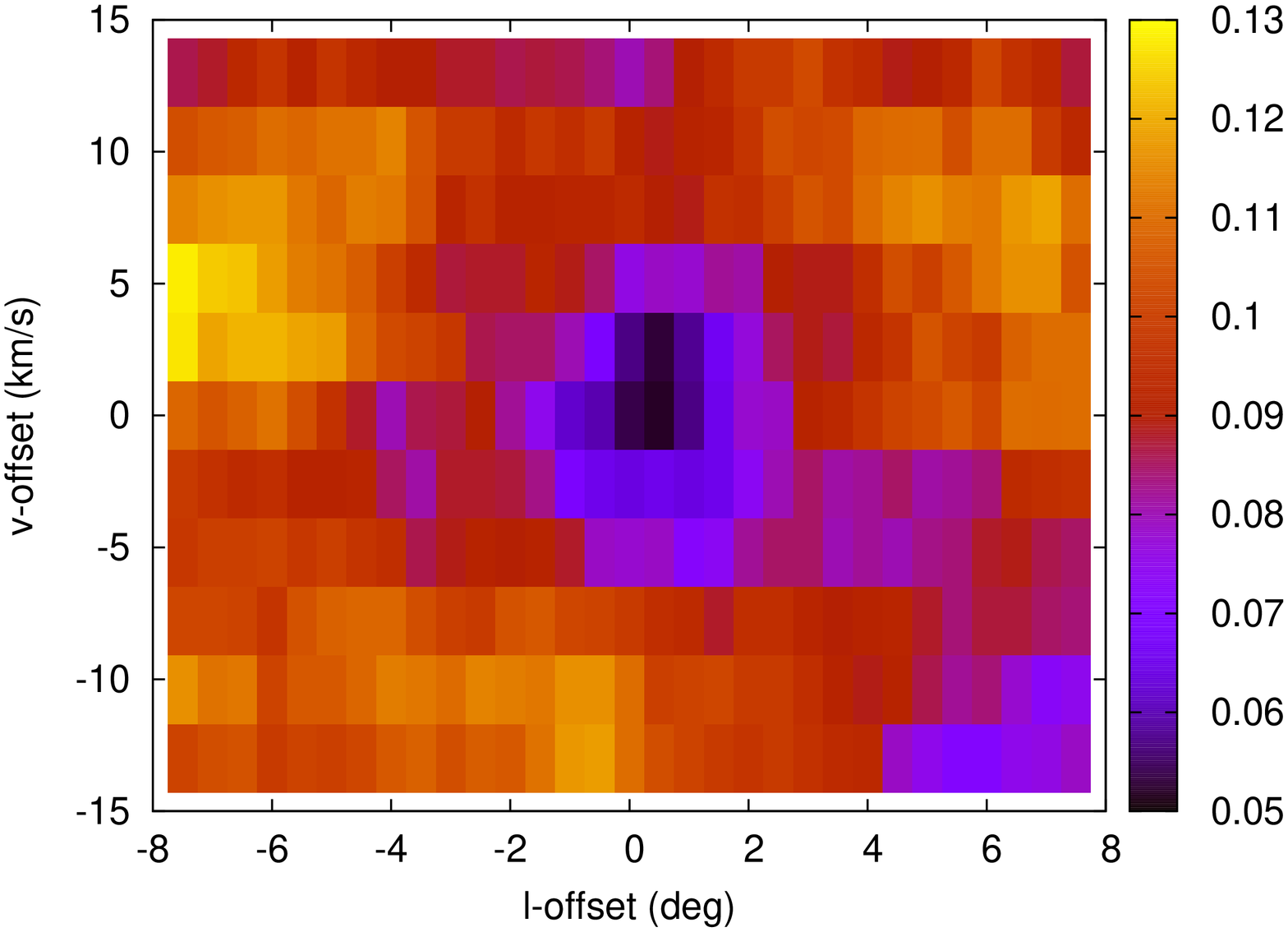} 
\includegraphics[angle=-90,width=0.45\linewidth]{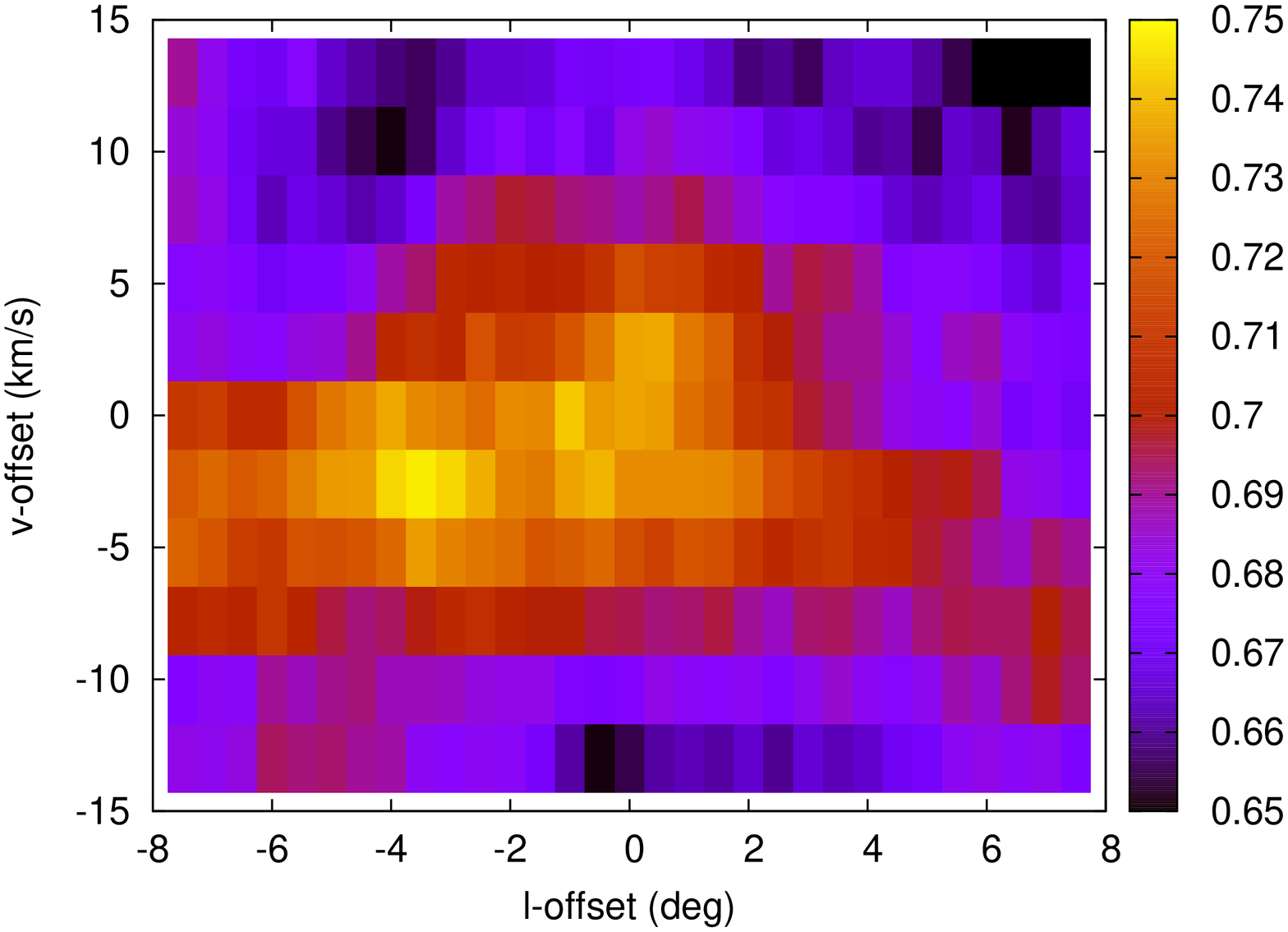} 
    \caption{Fraction of CO clumps falling into the interiors
of HI shells (left) or into their walls (right) as a function
of the artificial offset of CO clumps positions in l and v. 
Zero offsets correspond to the `real situation', non-zero offsets
test the robustness of the relation between HI shells and
CO clumps.
} 
       \label{glob_coclumpsoffset} 
\end{figure*} 
 
\begin{table} 
\caption{Average properties of CO clumps based on their 
position towards HI shells. 
} 
\label{tab_clumps1}      
\centering                          
\begin{tabular}{|lr|rrrr|}        
\hline 
env & $N_{\mathrm{cl}}$ & $T_{\mathrm{cl,HI}}$ & $T_{\mathrm{cl,CO}}$ & 
$\varepsilon _{\mathrm{cl}}$ & $T_{\mathrm{in}}/T_{\mathrm{out}}$ \\ 
\hline 
0:{\tiny average} & 2617 & 67 & 1.08 & 0.65 & 1.08 \\ 
1:{\tiny inner bubble} & 38 & 45 & 1.13 & 0.74 & 0.97 \\ 
2:{\tiny outer bubble} & 107 & 43 & 1.18 & 0.75 & 1.01 \\ 
3:{\tiny inner wall} & 1100 & 61 & 1.05 & 0.67 & 1.08 \\ 
4:{\tiny outer wall} & 820 & 70 & 1.09 & 0.64 & 1.08 \\ 
5:{\tiny outside} & 442 & 80 & 1.12 & 0.61 & 1.08 \\ 
6:{\tiny outside} & 104 & 85 & 1.11 & 0.59 & 1.10 \\ 
7:{\tiny outside} & 6 & 71 & 0.91 & 0.62 & 1.10 \\ 
\hline 
\end{tabular} 
\tablefoot{ 
The left panel gives the number of the environment and the number 
of CO clumps that reside there. The right panel gives the average 
properties of CO clumps: HI and CO temperatures, molecularization, and the ratio 
of HI temperatures in the clump and outside it. 
} 
\end{table} 
 
For the centre of each CO clump we find the relative distance  
$r_{\mathrm{rel}}$ to its nearest bubble using 
(Eq. \ref{eq:rreldef}) and then calculate the distribution of CO 
clumps as a function of 
$r_{\mathrm{rel}}$: $n_{\mathrm{cl}}(r_{\mathrm{rel}})$. 
Since the different $r_{\mathrm{rel}}$ are contained differently in the 
studied datacube, we compute the relative frequency of pixels with 
$r_{\mathrm{rel}}$ in certain interval of values:  
$q_{\mathrm{rrel}}(r_{\mathrm{rel}})$. 
With it we derive the normalized frequency of clumps 
$n_{\mathrm{cl, norm}}(r_{\mathrm{rel}})$: 
\begin{equation} 
{n_{\mathrm{cl, norm}} (r_{\mathrm{rel}})} = 
{{n_{\mathrm{cl}} (r_{\mathrm{rel}})} \over {q_{\mathrm{rrel}}(r_{\mathrm{rel}})}} 
\label{eq:nclnormdef} 
.\end{equation} 
 
Fig. \ref{cl_hist_rrel} shows the distribution of  
$n_{\mathrm{cl}}(r_{\mathrm{rel}})$, 
$f(r_{\mathrm{rel}}),$ and $n_{\mathrm{cl,norm}}(r_{\mathrm{rel}})$. 
The function $n_{\mathrm{cl,norm}}(r_{\mathrm{rel}})$ has a sharp maximum 
at $r_{\mathrm{rel}} = 0.0$ corresponding to walls of HI shells or, more 
precisely, to the transition between the low-density outer bubbles  
and high-density inner walls of HI shells. 
 
Table \ref{tab_clumps1} shows average properties of CO clumps in 
different media. Though the trends seen in Table \ref{tab_clumps1} 
are the same as in Table \ref{tab_env1}, the numbers are slightly 
different  because now the average is done in two steps, 
first for each clump  and then for all clumps, not in one step for 
all pixels as was done previously. Basically, in the first approach -- Table 
\ref{tab_env1} -- all pixels have the same weight, while in the second 
approach -- Table \ref{tab_clumps1} -- the weight depends inversely 
on the size of the clump to which the pixel belongs. 
Table \ref{tab_clumps1}, similarly to Table \ref{tab_env1}, depends 
on the temperature cutoffs,  but again, trends indicated in this Table
are not dependent on its precise value: the change from 0.3 K to 0.6 K  
does not give different results.

To test if the distribution of CO clumps in a relation to HI shells is random or not, 
we artificially shift positions of CO clumps by an offset in the l, b, or v direction
and then look at the number of clumps that fall to the bubbles, i.e. interiors
of HI shells --- environments 1 and 2) and to the walls (environments 3 and 4).
Offsets are mutually independent and are the same for all clumps.
Results are shown in Fig. \ref{glob_coclumpsoffset}. We only show
calculations with l and v offsets, since the b offsets can only be
small (our datacube is quite narrow in the b-direction). The left panel
of Fig. \ref{glob_coclumpsoffset}  shows the 
ratio between the number of CO clumps falling into the interiors 
of HI shells and the total number of clumps
as a function of the offset in l and v. This ratio has 
a clear minimum for the zero offsets, corresponding to the `real situation'. 
The right panel of Fig. \ref{glob_coclumpsoffset}
shows the fraction of CO clumps falling into walls of HI
shells. The largest values of this ratio are for the 
smallest offsets. These results show that CO clumps are not distributed randomly, but
rather in such a way that interiors of HI shells contain very
few CO clumps, while walls contain a lot of them.

\subsection{Close environment of CO clumps} 
 
\begin{figure} 
\centering 
\includegraphics[angle=-90,width=0.9\linewidth]{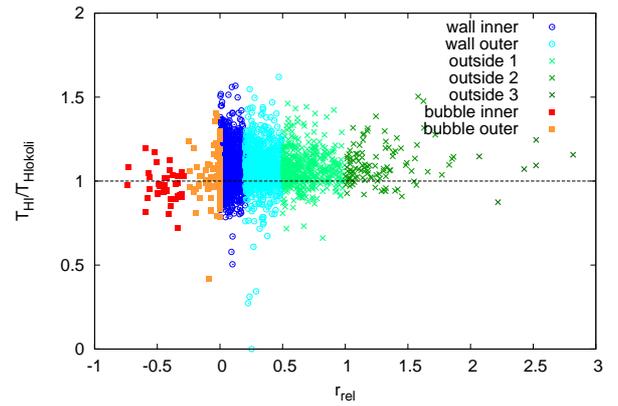} 
    \caption{Ratio of the inside and outside HI temperatures for 
CO clumps as a function of the relative distance $r_{\mathrm{rel}}$.} 
       \label{rrel_inout} 
\end{figure} 
 
For each CO clump we calculate the ratio of the average HI temperatures 
inside $T_{\mathrm{cl,HI}}$ and outside of it $T_{\mathrm{cl,HIout}}$ (for 
the calculation we used the outside as three times larger than the size 
of the clump, but results are similar for other values). In  
Fig. (\ref{rrel_inout}) 
we see that the spread in values is high, but that the clumps inside 
bubbles tend to have the ratio between inside and outside HI temperatures 
close to 1, while other clumps (dominated by wall clumps) 
tend to have a  slightly higher value of 1.1 (see average values in 
Table \ref{tab_clumps1}) 
Their dispersion seems to be much higher, but since the 
number of bubble clumps is small, we do not quantify it.

\subsection{Which HI shells support or inhibit CO creation} 
 
We tried to find a correlation between properties of HI shells 
and a number of associated clumps.  Unfortunately, we are 
very much bound by the relatively low resolution of observations. 
We found  that the larger the angular size of the shell or its velocity 
extent is, the more clumps it is associated with, which is 
obviously an effect of the resolution. The other correlation 
we found is  that the larger galactocentric distance of the HI shell  
(computed as the kinematic distance using shell radial velocity  
and the rotation curve of the Milky Way, see Paper 1) 
means fewer associated CO clumps, which is also not surprising. 
 
\begin{figure*} 
\centering 
\includegraphics[angle=-90,width=0.45\linewidth]{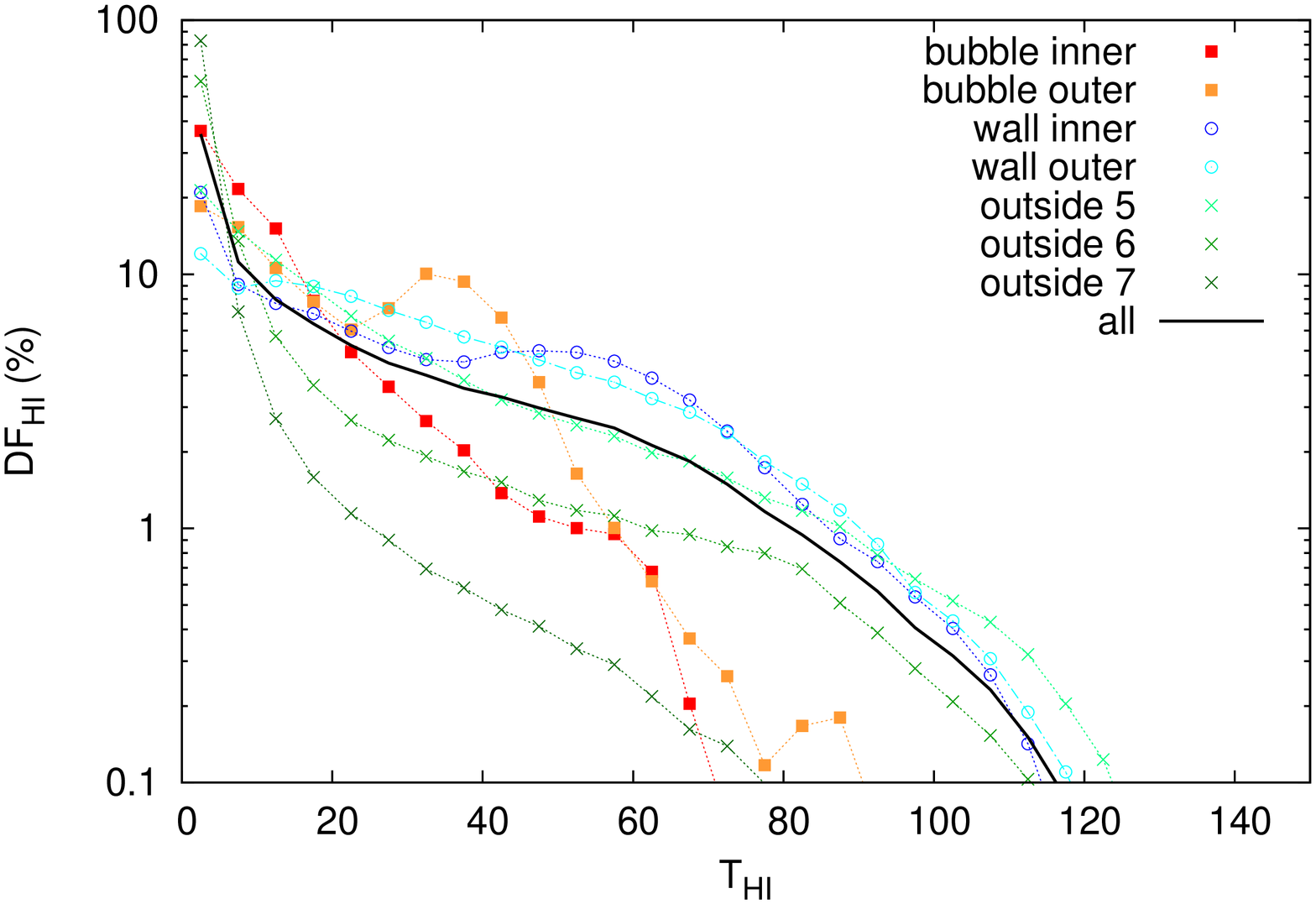} 
\includegraphics[angle=-90,width=0.45\linewidth]{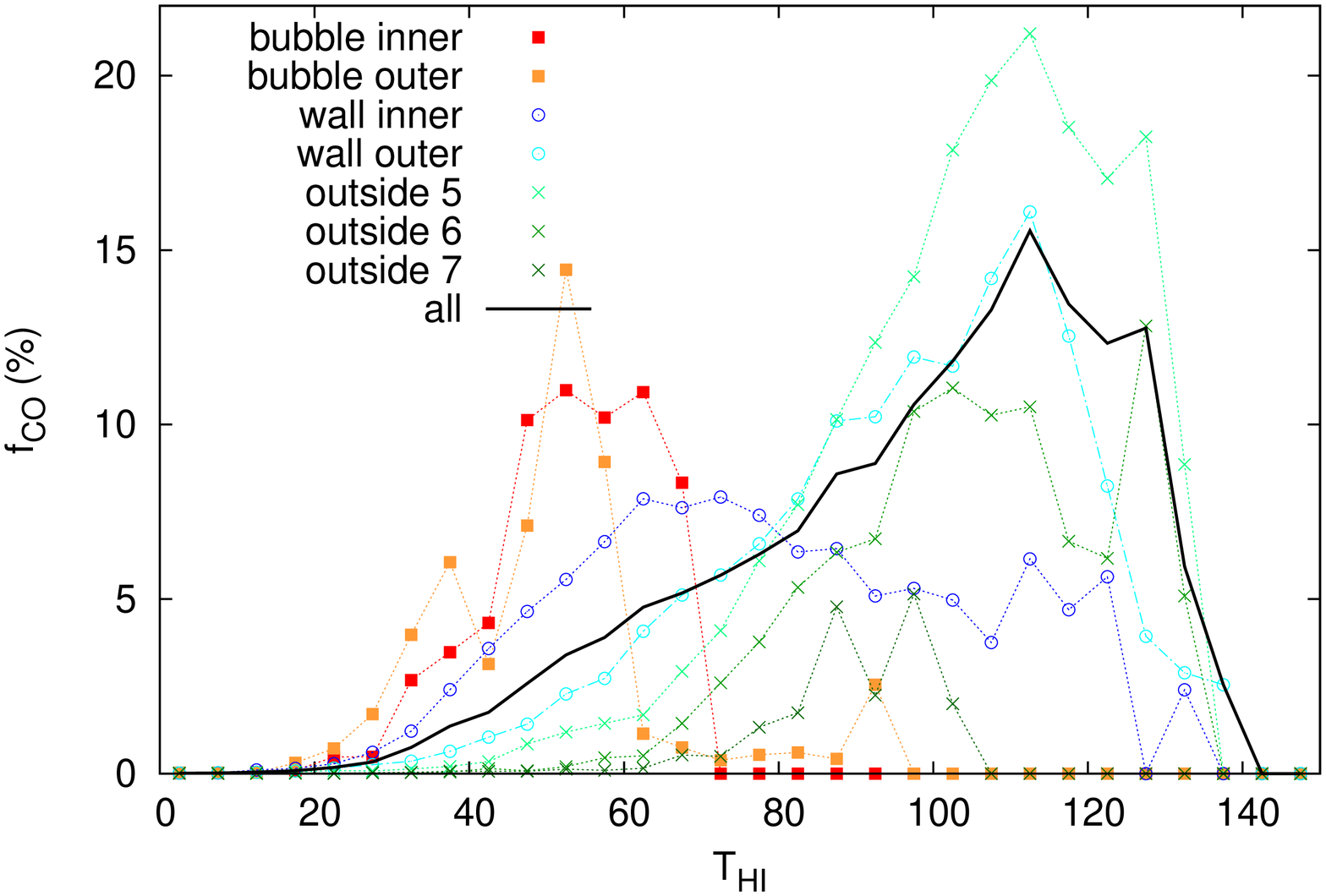} 
    \caption{Frequency of different HI temperatures (left) and the 
CO filling factor (right) as a function of the HI temperature.} 
       \label{tempdist1} 
\end{figure*} 
 
\begin{figure*} 
\centering 
\includegraphics[angle=-90,width=0.45\linewidth]{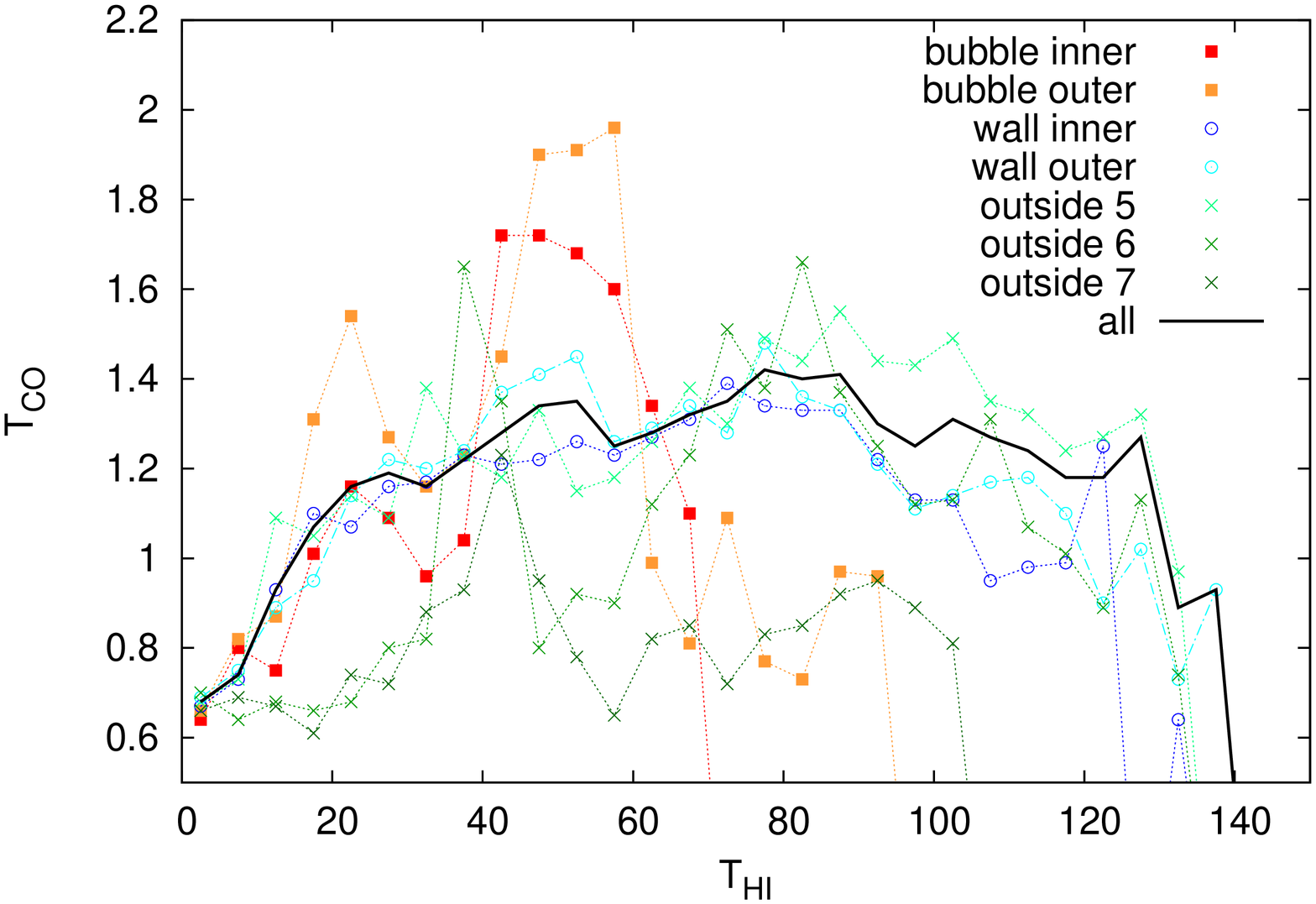} 
\includegraphics[angle=-90,width=0.45\linewidth]{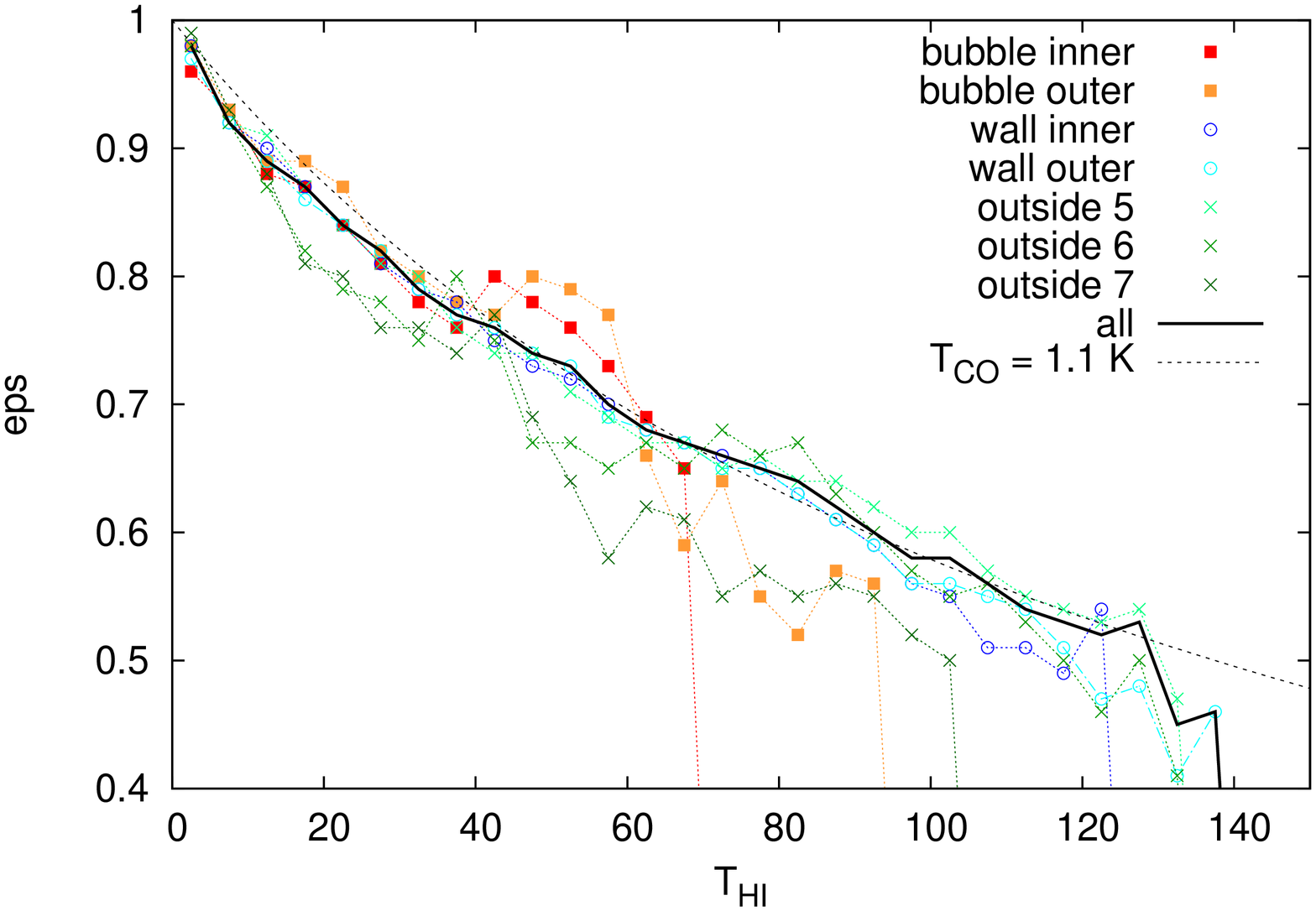} 
    \caption{Average CO temperature (left) and the 
molecularization (right) as a function of the HI temperature.} 
       \label{tempdist2} 
\end{figure*} 
 
\section{Discussion} 
 
\begin{figure} 
\centering 
\includegraphics[angle=-90,width=0.9\linewidth]{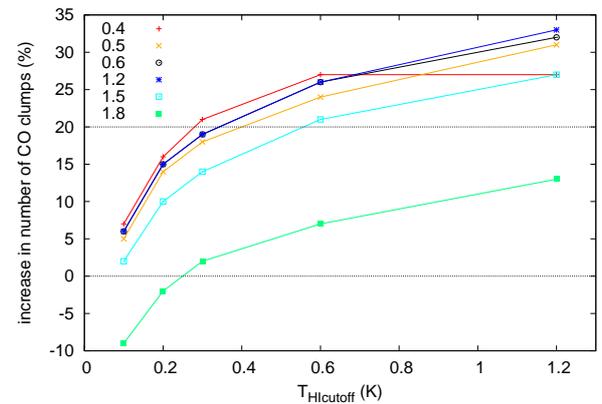} 
    \caption{Estimated increase in the number of CO clumps as  
     a function of CO and HI cutoffs. The x-axis gives the HI cutoff, 
     the y-axis shows the percent  increase in the number of clumps. 
     Different lines shows different CO cutoffs (cutoff values are shown in K). 
} 
       \label{fig_increase} 
\end{figure} 
 
\subsection{Rearrangement, destruction and creation of CO clumps} 
 
There is a real lack of CO clumps inside HI shells. And those 
that are inside have properties different from those in walls. 
Therefore, we assume that some of the preexisting CO clumps were 
destroyed by the activity that created HI shells;  a part survived, 
but their diffuse component (mostly HI) was removed; and a part 
was shifted to walls, away from the destructive forces. 
 
We find quite a strong evidence that preexisting CO clumps inside 
HI shells were changed: lower 
$T_{\mathrm{cl,HI}}$ , larger $T_{\mathrm{cl,CO}}$, 
larger $f_{\mathrm{CO}}$, lower $T_{\mathrm{cl,HI}}/T_{\mathrm{cl,HIout}}$. 
The removal of the diffuse component (HI and partly CO) seems 
to be a good explanation of these observations. 
What we do not know is what fraction of preexisting clumps was 
destroyed and what fraction was moved to the walls. Looking at 
Fig. \ref{tempdist1} we are inclined to say, that the inner wall 
contains a significant amount of pushed and robbed CO clumps, 
apart from newly created or preexisting mostly uninfluenced 
CO clumps, while the outer wall contains mainly the latter 
type of clumps. The HI temperature distribution  
$(DF_{\mathrm{HI}}$) 
is similar 
for both walls, but the dependence of the CO filling factor 
$f_{\mathrm{CO}}$ on the HI temperature differs and for the inner 
wall it looks like an intermediate stage between the profile 
for bubble CO clumps and outer wall clumps. 
 
It is easier to calculate the total outcome of these two 
conflicting effects, the destruction of preexisting clumps, and 
creation of new ones. 
We assume that the environment outside (5,6, and 7 in the tables) 
accurately describes  conditions that  are not influenced by the presence 
of HI shells. Then, using the values in Table \ref{tab_clumps1} we 
calculate that the total number of CO clumps in the studied region 
should be 1038. That is 2.5 times smaller than the measured value 
(which is 2617). 
 
This is slightly unfair towards the outside since this 
environment contains a lot of low-density (i.e. low HI temperature) 
gas, and the amount of CO depends on the density of ambient gas. 
If we restrict ourselves to the environment `outside 5', which 
has a similar HI temperature distribution for both walls (inner and outer) 
-- see $DF_{\mathrm{HI}}$ at Fig. \ref{tempdist1} -- and take these 
values as typical of the 
unperturbed medium, we come to the expected total number of CO clumps 
of 2210. The real value is larger by about 20 \%. 
 
The derived increase in the number of CO clumps depends on 
values in Tables \ref{tab_env1} and \ref{tab_clumps1}, which 
depend on temperature cutoffs $T_{\mathrm{COcutoff}}$ and 
$T_{\mathrm{HIcutoff}}$. Fig. \ref{fig_increase} shows the percent 
increase in the number of CO clumps as a function of these 
cutoffs. As canonical values we use  
$T_{\mathrm{HIcutoff}} = 0.3\ K$, corresponding to three times 
the rms noise in the data, and 
$T_{\mathrm{COcutoff}} = 0.3\ K$, which is equal to three times the 
rms noise, as explained in  section 2.  
The percent increase is shown, as the number of  
CO clumps depends very strongly on the $T_{\mathrm{COcutoff}}$ 
temperature, ranging from $\sim$ 3700 for  
$T_{\mathrm{COcutoff}} = 0.4\ K$, through $\sim$ 2600 for 
$T_{\mathrm{COcutoff}} = 0.6\ K$, to $\sim$ 1000 for 
$T_{\mathrm{COcutoff}} = 1.5\ K$.  
In Fig. \ref{fig_increase} we see that for CO cutoffs 
below $\sim 1.5\ K$, the increase is nearly independent  
of the value of the CO cutoff. Above 1.5 K, results  
substantially change, but then, values above 1.5\ K are 
probably too high to be used as a threshold. It is nonetheless 
interesting to see the difference in the behaviour of  
lower and higher density (corresponding to lower and higher 
brightness temperature) molecular gas in connection 
to HI shells.  
 
The increase in the number of CO clumps also depends   
on the HI cutoff. For reasonable values 
of this $T_{\mathrm{HIcutoff}}$, i.e. well above the rms noise in the 
HI data (0.09 K) and not too much above it, the resulting 
increase lies in the range between 15 and 25 \%.  
 
This 20 \% increase in the number of CO clumps is the net outcome 
where destroyed CO clumps are deducted from newly created CO clumps. 
It is a very rough estimate, but 
it seems that after all, HI shells (and more precisely, energetic 
activities that create HI shells) have on average a positive outcome 
on the amount of CO in the Galaxy. 
 
\subsection{Reliability of results} 
 
Our results are, obviously, dependent on what we consider to be HI 
shells and this is a source of two types of errors: 
\begin{itemize} 
\item{Problem 1: the dimensions of the shells are wrong (under- or 
overestimated); } 
\item{ Problem 2: an HI shell is not real, i.e. the identification 
is false.} 
\end{itemize} 
 
We deal with these problems separately. Owing to our 
way of calculating the relative distance to the nearest 
pixel of the bubble, and to our clean separation of bubble 
and other pixels, Problem 1 is not so serious. It leads 
to slightly incorrect relative distances. Pixels close to  
the interesting bubble/wall frontier are not  influenced so much. This  
problem leads 
to `smearing' in observed behaviour, but not to different trends. 
 
Problem 2 is more serious and its influence cannot be so easily  
estimated. In this paper we take as a basis for our analysis 
the catalogue of HI shells from Paper I. This catalogue -- like other  
catalogues of HI shells -- probably contains some false identifications  
and  probably misses some real structures as well.  
We discuss the different identifications in Paper I where a comparison  
between different catalogues of HI shells is provided. A similar comparison  
is given in \citet[][]{2014A&A...564A.116S},  
which is also a catalogue of HI shells in the outer Milky Way, 
based on the LAB HI survey, but using a very different identification 
technique. Our identification technique is based on locating 
HI holes, while theirs is based on locating HI walls. In 
an ideal case these two approaches should find the same 
objects, but we can easily imagine (and find in datacubes) structures that do not have easily 
identifiable walls or, on the other hand,  
structures that are formed mostly by fragmented walls. 
 
The main result of the comparison in \citet[][]{2014A&A...564A.116S} is 
that there are shells that  are found  in only one of these 
catalogues: Suad's catalogue contains about twice as many shells as ours 
since they also include open structures, but they reidentify  
slightly less than half of our structures. 
More recently, \citet[][]{2015AJ....149..189S} 
give a comparison of shell catalogues based on observations with  
different angular resolutions.  
 
To understand if and how much we can trust results based on properties of  
shells in our catalogue, when we know 
that our catalogue is not perfect, we constructed a toy model:  
it consists of $N_{\mathrm{tot}}$ cubes, the cube 
can be either empty (i.e. filled by the unperturbed medium)  
or contain the shell (i.e. bubble and wall). 
Each of these environments  (unperturbed, bubble, and wall) has the  
value of a hypothetical quantity $q$. 
 
Now, we assume that by using an identification algorithm some  
of these existing structures are identified 
as structures (their number is $N_{\mathrm{detected}}$), but some 
are missed ($N_{\mathrm{missed}}$). On the other hand, some empty cubes 
are falsely identified as structures ($N_{\mathrm{false}}$). With these  
identifications containing errors we calculate observed values of  
$q$ and compare them 
to real values and look, how much they are changed by missing and false 
identifications. 
 
To mimic an analysis we performed in this paper, we define  
the values of the quantity $q$ as follows: in the unperturbed medium 
 $q_{\mathrm{unperturbed}} = 1.0$, in the walls 
$q_{\mathrm{wall}} = 2.0$, and inside the bubble $q_{\mathrm{bubble}} = 0.0$.  
Then, we derive observed values of $q$ from our  
identifications with errors.  
If $q_{\mathrm{obs,wall}} \geq 1.5$ and  
$q_{\mathrm{obs,bubble}} \leq 0.5$, we consider the experiment 
 successful, otherwise it is considered  marred 
by incorrect identifications. Filling factors of bubble and wall 
environments were chosen to resemble those in Table \ref{tab_env1}, 
but they were found not to influence following results. 
 
We define the `badness' of the experiment as 
\begin{equation} 
\mathrm{badness} =  
{{N_{\mathrm{false}} + N_{\mathrm{missed}}} \over 
{N_{\mathrm{detected}}}} 
\label{eq:toy_bad} 
\end{equation} 
and porosity as 
\begin{equation} 
\mathrm{porosity} =  
{{N_{\mathrm{detected}} + N_{\mathrm{missed}}} \over 
{N_{\mathrm{detected}}}} 
\label{eq:toy_poro} 
\end{equation} 
and we found, that the experiment is successful, if 
\begin{equation} 
{{\mathrm{badness}} \over {\mathrm{porosity}}} = 
{{N_{\mathrm{false}} + N_{\mathrm{missed}}} \over 
{{N_{\mathrm{detected}} + N_{\mathrm{missed}}}}} \leq 1.0 
\label{eq:toy_badporo} 
\end{equation} 
 
Observed values of $q_{\mathrm{obs,unperturbed}}$ usually 
differed from the input value $q_{\mathrm{unperturbed}}$, but for successful 
experiments it was always in between  
$q_{\mathrm{obs,bub}} < q_{\mathrm{obs,unperturbed}} < q_{\mathrm{obs,wall}}$. 
It does not depend on some simple criterion like condition \ref{eq:toy_badporo}. 
 
 In the case of nearly no 
missed structures condition \ref{eq:toy_badporo} leads to  
$N_{\mathrm{false}}/N_{\mathrm{detected}} \leq 1.0$, i.e. to a quite 
reassuring condition that if at least half of the structures  
in the catalogue are real, then we might get good enough  
results. Taking values  
$N_{\mathrm{false}} \sim 0.5 N_{\mathrm{detected}}$ 
and  $N_{\mathrm{missed}} \sim N_{\mathrm{detected}}$ 
(which are based on the comparison of catalogues  
in \citet[][]{2014A&A...564A.116S}) as typical values for 
our  HI shells catalogue (and perhaps any catalogue), we get from  
Eq. \ref{eq:toy_badporo} the result 
$\mathrm{badness}/\mathrm{porosity} \sim 0.5$, which means 
success, i.e. that the  observed trends are real. 
 
Our toy model is certainly very simple, but at least we can tell 
that our analysis in this paper might detect real trends in the 
Galaxy, even though there are imperfections in our input data, 
especially in the catalogue of HI shells. 
 
\section{Summary}

The distribution of CO clumps is correlated with HI shells: the  
interiors of bubbles are devoid of both HI and CO. 
CO clumps tend to sit in the walls of HI shells (Fig. \ref{glob_rrel1}), 
they are concentrated in the region between low HI temperature 
bubbles and high HI temperature walls of the shells. 
CO is more concentrated than HI to the bubble/wall transitions. HI 
walls are much thicker; the typical thickness is 
$\sim 0.5 r_{\mathrm{rel}}$. 
 
The increased molecularization in bubble CO clumps is mostly 
given by the lower values of HI temperatures (as a consequence 
of Eq. \ref{eq:epsdef}), though it might be slightly increased 
compared to wall CO clumps, as seen in Fig. \ref{tempdist2}. 
 
The HI environment of bubble CO clumps seems to be rather flat. 
There is no increase in HI inside the clumps and there are no 
big changes outside it (see Fig. \ref{rrel_inout} for the value 
and dispersion of the ratio inside and outside HI temperature). 
Actually, not even wall CO clumps coincide with the HI maxima 
(they probably coincide with past HI maxima), but the region around 
them is typically more diverse and 
less flat  than around bubble CO clumps. 
 
CO clumps inside bubbles have relatively low HI temperatures 
compared to CO clumps in walls or outside --- it is nicely 
seen in Fig. \ref{cl_tcothi}  --- 
but for this lower 
HI temperature they tend to have larger CO filling factor 
$f_{\mathrm{CO}}$ (see Fig. \ref{tempdist1}). 
 
\section{Conclusions} 
 
Based on our analysis of HI shells in the outer Milky Way, we draw the  
following conclusions: 
\begin{itemize} 
\item{In interiors of HI shells and in regions of the decreased HI density, 
 CO clumps are smaller and contain less HI than 
 clumps in walls of HI shells or clumps outside HI shells. It seems 
 that some CO clumps inside these bubbles were destroyed and the remaining 
 ones were robbed of their more diffuse gas.} 
\item{A substantial amount of CO clouds exist outside HI shells. 
 They are often quite massive and tend to have a large HI content. 
 They are probably formed independently of HI shells.}  
\item{In walls of HI shells there is an increased occurrence 
 of CO clumps. This increase cannot be simply explained by the 
 higher HI densities and looking at differences between clumps in 
 walls and in an unperturbed medium outside of HI shells, we  
 estimate the total increase of CO in the outer Milky Way due to 
 HI shells being around 20 \%.}   
\end{itemize}

\begin{acknowledgements} 
This study has been supported by the Czech Science 
Foundation grant 209/12/1795,  
the M\v{S}MT grant LG14013 and by the project RVO: 67985815.
The authors would like to thank  the anonymous referee for constructive 
comments leading to the improvement of the paper. 
\end{acknowledgements} 
 
\bibliographystyle{aa} 
\bibliography{ehlerova16_buhico_ref}

\end{document}